\documentclass[aps,prd,twocolumn,showpacs,superscriptaddress,nofootinbib,tightenlines,floatfix]{revtex4-2}

\usepackage{hyperref}
\usepackage{physics}
\usepackage{amsmath,amssymb}
\usepackage{graphicx,multirow}
\usepackage{davitex}
\hypersetup{
    colorlinks=true,     
    linkcolor=blue,      
    citecolor=blue,      
    filecolor=blue,      
    urlcolor=blue,        
    linktoc=page
}

\newcommand{\chimax}{\chi_{P,{\rm max}}}
\newcommand{\houston}{University of Houston, Houston, Texas 77204, USA}

\begin{document}

\title{Deep learning of phase transitions with minimal examples}

\author{Ahmed Abuali}\email{amabuali@uh.edu}
\affiliation{Physics Department, \houston}

\author{David A. Clarke}\email{clarke.davida@gmail.com}
\affiliation{Department of Physics and Astronomy, University of Utah, 
Salt Lake City, Utah 84112, USA}

\author{Morten Hjorth-Jensen}
\affiliation{Department of Physics and Astronomy and Facility for Rare Isotope Beams,\\
Michigan State University, East Lansing, Michigan 48824, USA}
\affiliation{Department of Physics and Center for Computing in Science Education, University of Oslo, N-0316 Oslo, Norway}

\author{Ioannis Konstantinidis}
\affiliation{Computer Science Department, \houston}

\author{Claudia Ratti}
\affiliation{Physics Department, \houston}

\author{Jianyi Yang}
\affiliation{Computer Science Department, \houston}

\date{\today}

\begin{abstract}
Over the past several years, there have been many studies demonstrating the ability
of deep neural networks
to identify phase transitions in many physical systems,
notably in classical statistical physics systems.
One often finds that the prediction of deep learning methods trained
on many ensembles below and above the critical temperature $\Tc$ behaves similarly
to an order parameter, and this analogy has been successfully used to 
locate $\Tc$ and estimate universal critical exponents.
In this work, we pay particular attention to the ability of a convolutional
neural network to capture these critical parameters for the 2-$d$ Ising model 
when the network is trained on configurations at $T=0$ and $T=\infty$ only.
We directly compare its output to the same network trained at
multiple temperatures below and above $\Tc$ to gain understanding of how this extreme restriction
of training data can impact a neural network's ability to classify phases.
We find that the network trained on two temperatures is still able to
identify $\Tc$ and $\nu$, while the extraction of $\gamma$ becomes more challenging.
\end{abstract}

\maketitle

\section{Introduction}

A change of phase can often be characterized by an order parameter, an observable 
whose value is zero in one phase and nonzero in the other.
In classical systems, when an order parameter can be
identified, we call this phase change a phase transition. For phase transitions,
there is a unique critical point or threshold in the control parameter at which
the logarithm of the partition function $\log Z$ is not analytic.
In other cases, there exist paths in
the space of control parameters along which the phase changes while
$\log Z$ experiences no non-analyticities. Such cases are called
crossovers. Crossovers have no order parameter, and identifying
a pseudocritical phase boundary, e.g. a pseudocritical temperature $\Tpc$, 
is inherently ambiguous, as different observables generally exhibit
pseudocritical behavior at different values of the
control parameters.

Locating such boundaries is of special interest to the phase diagram
of nuclear matter, whose change of phase from a gas of hadrons and their
resonances to quark-gluon plasma at zero net-baryon chemical potential
is known to be a crossover~\cite{Aoki:2006we}.
The most commonly used quantity to define $\Tpc$ is the light-quark
chiral condensate $\ev{\bar\psi_l \psi_l}$, which is the order parameter of
the chiral transition in the limit of zero light-quark mass, and observables 
deriving from $\ev{\bar\psi_l \psi_l}$ deliver $\Tpc\approx158$ 
MeV~\cite{HotQCD:2018pds,Borsanyi:2020fev}. This temperature roughly agrees with
$\Tpc$ taken from certain observables deriving from the Polyakov 
loop, the order parameter of the deconfinement transition in the
limit of infinite quark mass~\cite{Bazavov:2016uvm,DElia:2019iis}.
On the other hand, the Polyakov loop susceptibility itself exhibits its
peak only above $T\approx180$ MeV, implying a much higher $\Tpc$~\cite{Bazavov:2016uvm}.
Moreover, there is strong evidence that heavier hadronic excitations,
in particular charmed hadronic excitations, persist even up to 
$T\approx175$ MeV~\cite{Bazavov:2023xzm}.

Given the ambiguity of demarcating a critical threshold,
it may be illuminating to spot phase
changes without specifying a physically motivated observable ahead of time.
One approach is to use machine learning. 
Machine learning, especially neural networks (NN), has proven to be particularly 
well suited for phase classification tasks and locating critical points
in classical spin systems, for both supervised and unsupervised learning,
using a variety of architectures; see e.g.~\cite{Carrasquilla:2016oun,
Wang:2016nmn,
vanNieuwenburg:2016zsd,
Li:2017xaz,
Hu:2017wey,
Wetzel:2017olt,
Rodriguez-Nieva:2018cbl,
Foreman:2018ktj,
Cossu:2018pxj,
ShibaFunai:2018aaw,
Tan:2019eih,
Bachtis:2020dmf,
Walker:2020hiq,
Han:2022mhy,
Bae:2024,
Li:2025rbf}. 
For studies of classical spin systems that use NNs with supervised learning, 
one typically trains the
NN on multiple ensembles below and above the critical temperature $\Tc$.
These ensembles consist of configurations generated through Markov chain Monte Carlo (MCMC) simulations \cite{Gilks1995}.
Configurations for ensembles with $T<\Tc$ are labeled ``ferromagnetic", while
ensembles with $T>\Tc$ are labeled ``paramagnetic". 

In the context of a crossover, this
approach presents the inherent difficulty that there is no unique $\Tc$, and therefore
such a labeling is ambiguous in the region where the phase changes.
Moreover, training using many temperatures presents a significant
computational burden that grows with the number of examples.
To circumvent these issues without using unsupervised learning, we 
consider a convolutional neural network (CNN) \cite{Goodfellow2016}
trained only at $T=0$ and $T=\infty$, where the system is unambiguously ferromagnetic
and paramagnetic, respectively. 
Rather than generating these two ensembles
using MCMC simulations, we use the system's known infinite-volume characteristics to create
ensembles of exact magnetizations $|m|=1$ and $m=0$. 
This approach has the further advantage of significantly reducing the
computational effort required for training, which is especially important
when studying systems where each site
is associated with large numbers of continuous degrees of freedom.
A similar strategy was applied in Ref.~\cite{Li:2017xaz} to
a CNN with a variety of 2-$d$ Potts models with $q$ states from $q=2$ to $q=10$.
There, they trained only on each model's theoretical $T=0$ configurations and were
able to distinguish when a model's transition was of first or second order.

Here, we instead focus on a quantitative comparison of the capabilities of 
this type of approach to the capabilities
of a more conventional supervised learning approach that trains on multiple
ensembles around the transition region~\cite{Bachtis:2020dmf}. 
In particular, we focus on the 2-$d$ Ising model, which we choose because it
features a true phase transition, allowing us to calibrate each training
regimen using the model's analytically
known critical parameters. We examine the ability
of each regimen to accurately extract $\Tc$ and critical exponents,
to gain some understanding on what is ``lost" when using a minimal set
of examples for phase transition classification. While we were motivated
by the study of crossover phenomena, we leave detailed, quantitative analysis
of crossover systems to future work.

The outline of this paper is as follows. We start in \secref{sec:ising} by describing our statistical physics model along
with our general deep learning approach. In \secref{sec:models}, we describe our Monte
Carlo data and training method in detail. We show and discuss our findings in \secref{sec:results},
wrapping up in \secref{sec:summary} with a brief conclusion.

\section{Deep learning and the Ising model}\label{sec:ising}

We use a CNN
to classify the phase of configurations belonging to the 2-$d$ Ising
model with zero external field, whose Hamiltonian is given by
\begin{equation}\label{eq:ising}
H=-J\sum_{\ev{i j}} \sigma_i \sigma_j,
\end{equation}
where the brackets indicate a sum over nearest neighbors,
$\sigma_i\in\{+1,-1\}$ is the spin at site $i$, and $J$ is the
nearest-neighbor interaction strength. We work
in units with $J=k_B=1$ and on square lattices of size $V=L^2$.
The order parameter is the magnetization $m=V^{-1}\sum_i\sigma_i$.
In the thermodynamic limit $L\to\infty$, 
the 2-$d$ Ising model is analytically known~\cite{Onsager:1943jn} to exhibit a
second-order transition at a critical temperature
$\Tc\approx 2.269185$, 
at which the magnetic susceptibility 
\begin{equation}\label{eq:magSusc}
\chi= \beta V\left(\ev{m^2}-\ev{m}^2\right),
\end{equation}
where $\beta=1/T$,
diverges. When $L<\infty$, $\chi$ peaks at a pseudocritical temperature
$\Tc(L)$. The approach of $\Tc(L)$ to $\Tc$ as $L\to\infty$ is controlled by
the universal critical exponent $\nu$. In particular,
\begin{equation}\label{eq:nuTc}
|\Tc(L)-\Tc|\sim L^{-1/\nu}.
\end{equation}
Meanwhile, the peak height $\chi_{\rm max}$ in this limit
scales according to the critical exponent 
$\gamma$ as
\begin{equation}\label{eq:gamma}
\chi_{\rm max}\sim L^{\gamma/\nu}.
\end{equation}
For the 2-$d$ Ising universality class, these exponents are
$\nu=1$ and $\gamma=1.75$~\cite{Onsager:1943jn}.
For discussions of these parameters, see for example Refs.~\cite{Cardy1996,Stanley1971,Newman1999}.

The pioneering study in Ref.~\cite{Carrasquilla:2016oun} trained a neural network  
to determine whether 2-$d$ Ising-model configurations belong to the ferromagnetic
or paramagnetic phase. After training on 
a broad range of ensembles below and above $\Tc$,
the authors of \cite{Carrasquilla:2016oun}
present the NN with configurations at new temperatures;
the output layer or prediction of the model $P$ 
is constructed to indicate the likelihood that a configuration is ferromagnetic,
with $P=1$ indicating a completely ferromagnetic phase and $P = 0$ indicating a completely paramagnetic phase. 
They noticed that the average prediction $\ev{P}$
behaves analogously to $\ev{m}$ and hence used $\ev{P}$ to extract numerical values 
of $\Tc$ and $\nu$ to good accuracy.
This analogy has been exploited in later studies, for example
Ref.~\cite{Bachtis:2020dmf}, 
which found good results also for $\gamma$.
In such contexts, $\ev{P}$ is sometimes thought of as an ``effective order parameter",
and one can construct in analogy to Eq.~\eqref{eq:magSusc} a
prediction susceptibility 
\begin{equation}\label{eq:predictionSusc}
\chi_P\equiv \beta V\left(\ev{P^2}-\ev{P}^2\right).
\end{equation}

In Ref.~\cite{Bachtis:2020dmf} it was noted that since $P$ is a function
of the configuration, it can be thought of as a thermodynamic observable and
therefore can be reweighted~\cite{Ferrenberg:1988yz,Ferrenberg:1989ui}. When reweighting,
one infers the average value of an observable $X$ at some $\beta'$ from
the known value at a sufficiently nearby point $\beta$ through
\begin{equation}\label{eq:RW}
  \ev{X}_{\beta'}
                 =\ev{\frac{Z_\beta}{Z_{\beta'}}e^{(\beta-\beta')H}X}_\beta,
\end{equation}
where $Z_\beta$ is the system's partition function at $\beta$.
We found reweighting to be useful in our context to accurately estimate
the location of the peak in the prediction susceptibility and 
estimate systematic uncertainties.

\begin{figure*}
\centering
\includegraphics[width=0.51\linewidth]{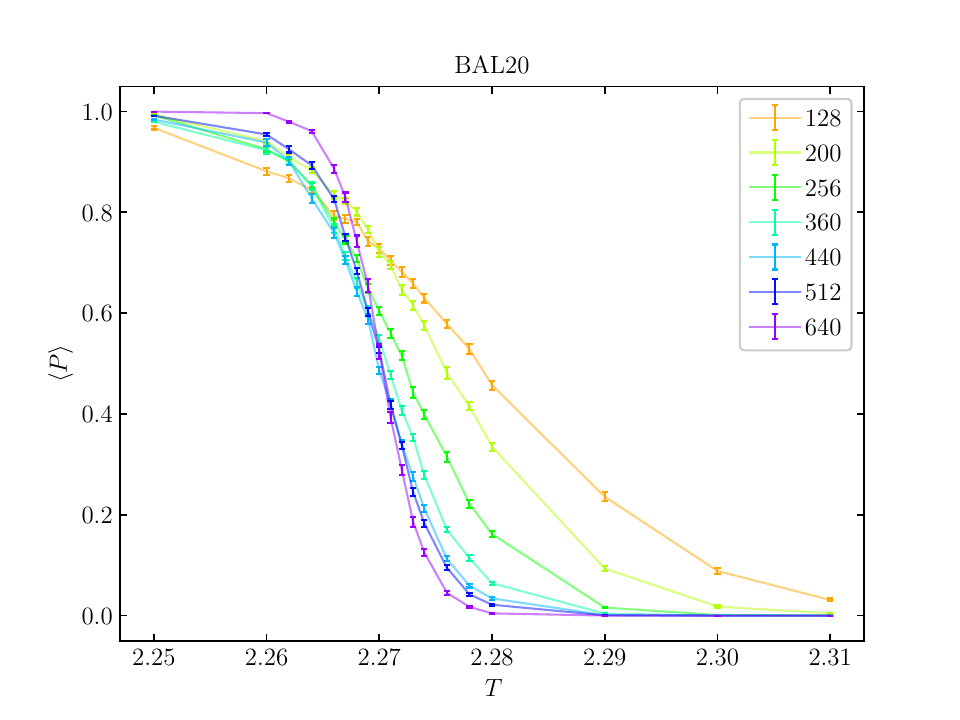}\hspace{-7mm}
\includegraphics[width=0.51\linewidth]{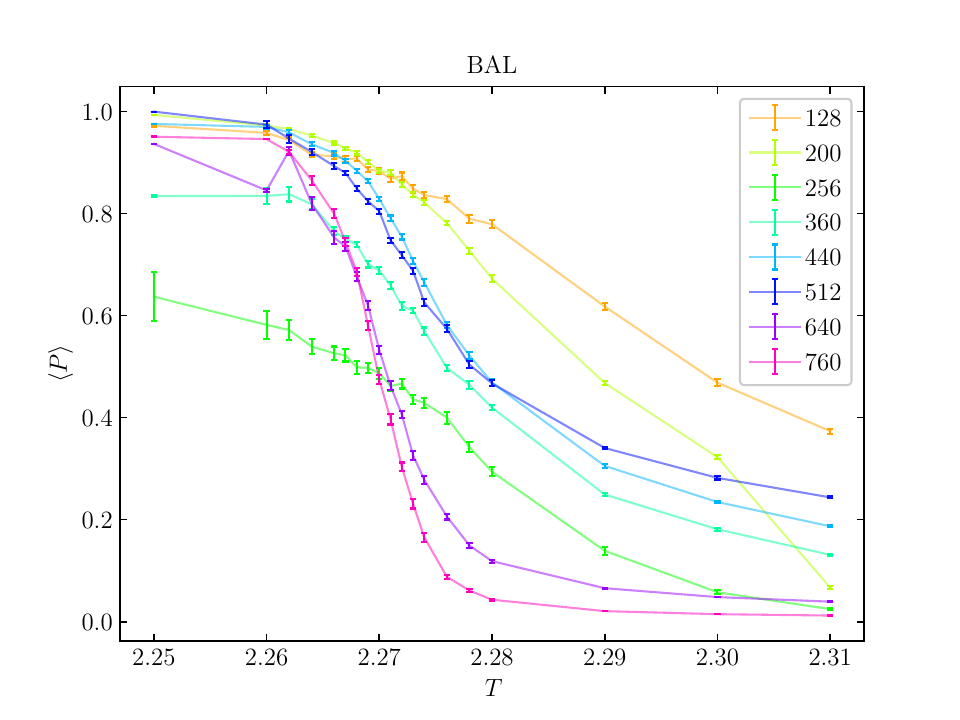}
\caption{$\ev{P}$ for the BAL20 model (left) and BAL model (right)
for each lattice size. Lines are drawn to guide the eye.}
\label{fig:PwithL}
\end{figure*}

\section{Set up and Deep Learning model}\label{sec:models}

We employ two training strategies in this work. The architecture we use is
exactly the same as the CNN given in Appendix B of Ref.~\cite{Bachtis:2020dmf}.
We choose this architecture as it was shown there to accurately find
$\Tc$, $\nu$, and $\gamma/\nu$.
Following Ref.~\cite{Bachtis:2020dmf}, our first strategy trains the 
CNN on 20 temperatures chosen above and below $\Tc$,
10 in the range $2.44\lesssim T\lesssim 3.13$ and
10 in the range $1.79\lesssim T\lesssim 2.23$,
corresponding approximately to the $\beta$ values listed in that work.
We call this the ``BAL20" model. 
Our second strategy trains the CNN on two ensembles only, 
corresponding to $T=0$ and $T=\infty$. The $T=0$ ensemble consists of 1000
configurations with $m=1$ exactly and 1000 configurations with $m=-1$.
Meanwhile, the $T=\infty$ ensemble consists of 2000 configurations
where a randomly chosen set of half the spins is set to $-1$
while the other half is set to $+1$, yielding $m=0$. We call this
the ``BAL" model. We emphasize that, in contrast to the BAL20
model, the BAL model is not trained on MCMC data. 
Our implementation of the CNN utilizes the machine-learning libraries Scikit-Learn \cite{sklearn} and Tensorflow \cite{tensorflow} with Keras \cite{keras}. This implementation
can be accessed publicly on GitHub~\cite{CNNGIT}.

\begin{figure*}
\centering
\includegraphics[width=0.51\linewidth]{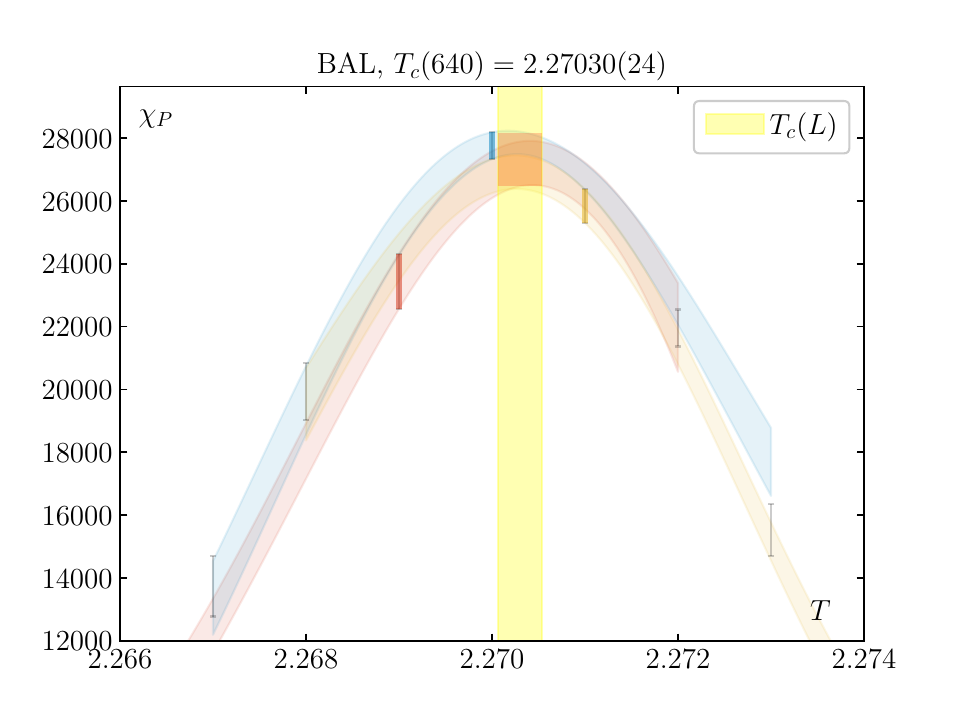}\hspace{-7mm}
\includegraphics[width=0.51\linewidth]{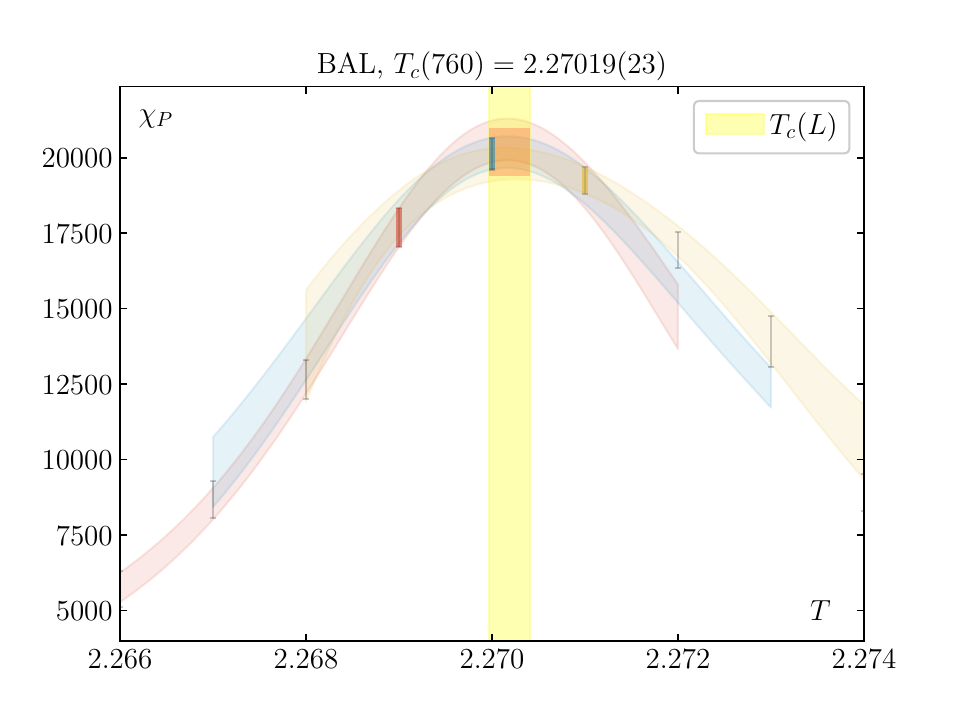}\\[-2mm]
\includegraphics[width=0.51\linewidth]{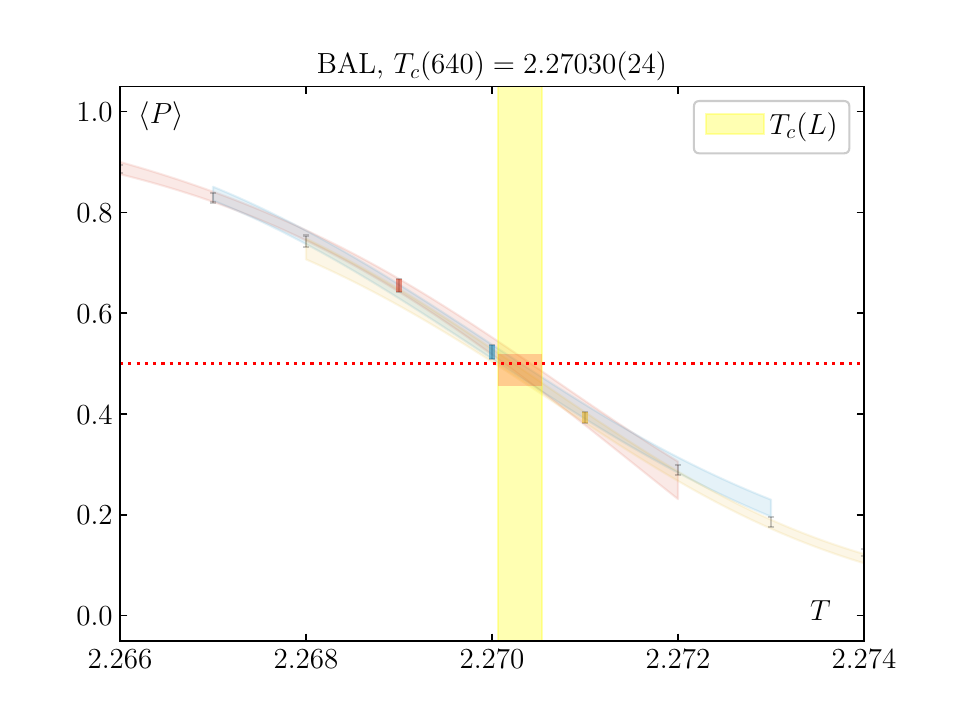}\hspace{-7mm}
\includegraphics[width=0.51\linewidth]{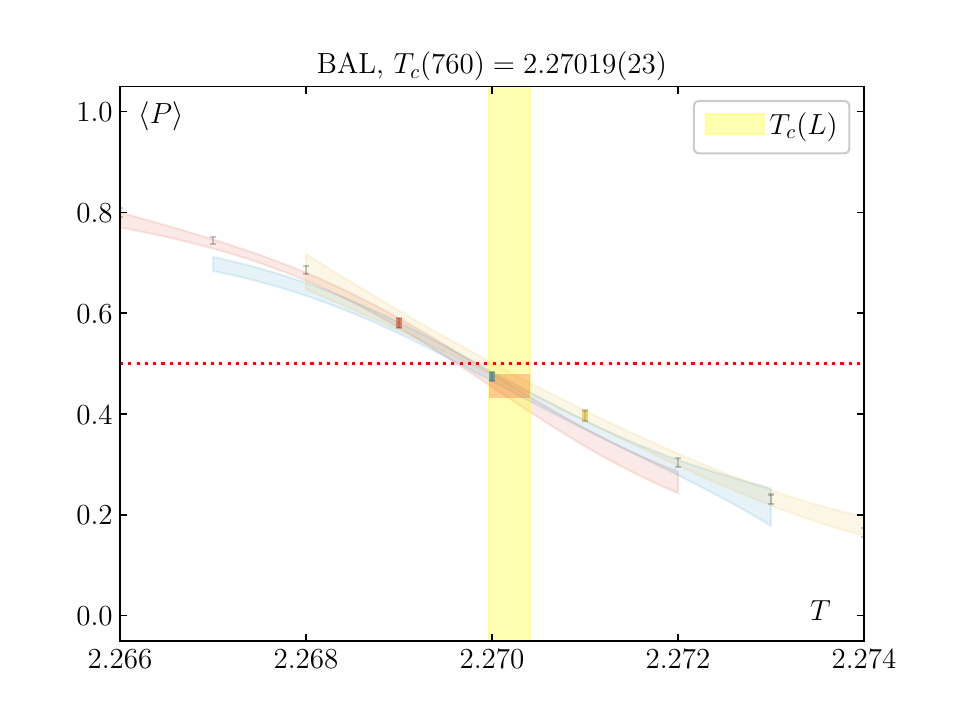}
\caption{Reweighting of $\chi_P$ (top) and $\ev{P}$ (bottom) for the 
BAL20 (left) and BAL (right) models using the largest lattice available to each
model. Data points indicate the output of the CNN. Each colored datum indicates
the starting point for a reweighting curve, which is
shown using the respective color. The yellow, vertical band 
indicates the estimated $\Tc(L)$. The orange box shows the reweighted observable
evaluated at $\Tc(L)$ along with its associated total uncertainty.
The red, dotted line indicates $\ev{P}=0.5$.}
\label{fig:RW}
\end{figure*}

For efficient generation of 2-$d$ Ising model configurations, we build on
the code of Ref.~\cite{isingcode}.
We generate configurations of size $V= L^2$ for $L=128$, 200, 256, 360, 440, 512, 
and 640. The BAL model has an additional set of ensembles
at $L=760$, as it is significantly cheaper to train.
For each $L$, we generate ensembles at many temperatures $T$.
For each $T$ we generate $\nconf=2000$ configurations that are extremely well
separated in Markov time; in particular, each measurement is separated
by $5L^2$ sweeps. Ensemble averages $\ev{X}$ are estimated by
\begin{equation}
\ev{X}=\frac{1}{\nconf}\sum_{i=1}^{\nconf} X_i.
\end{equation}
Statistical uncertainties are estimated using jackknife resampling with 40 bins.
Jackknife resampling, reweighting, fitting for the exponent
$\gamma$, and the subsequent model averaging discussed in \secref{sec:results} 
are carried out using software of the
AnalysisToolbox~\cite{Clarke:2023sfy}.

\begin{figure*}
\centering
\includegraphics[width=0.51\linewidth]{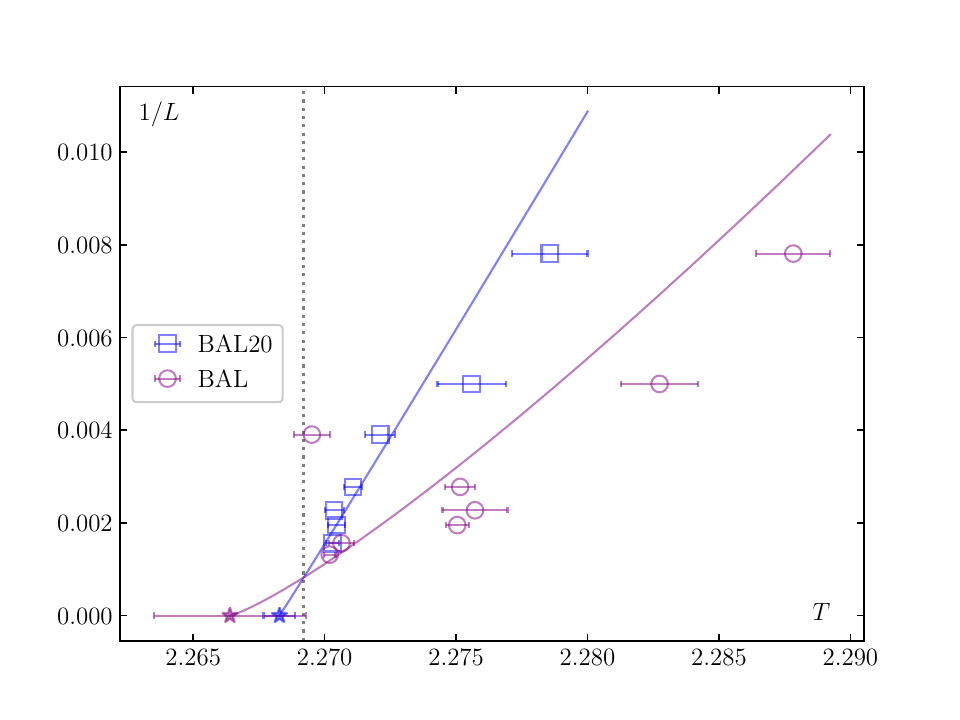}\hspace{-7mm}
\includegraphics[width=0.51\linewidth]{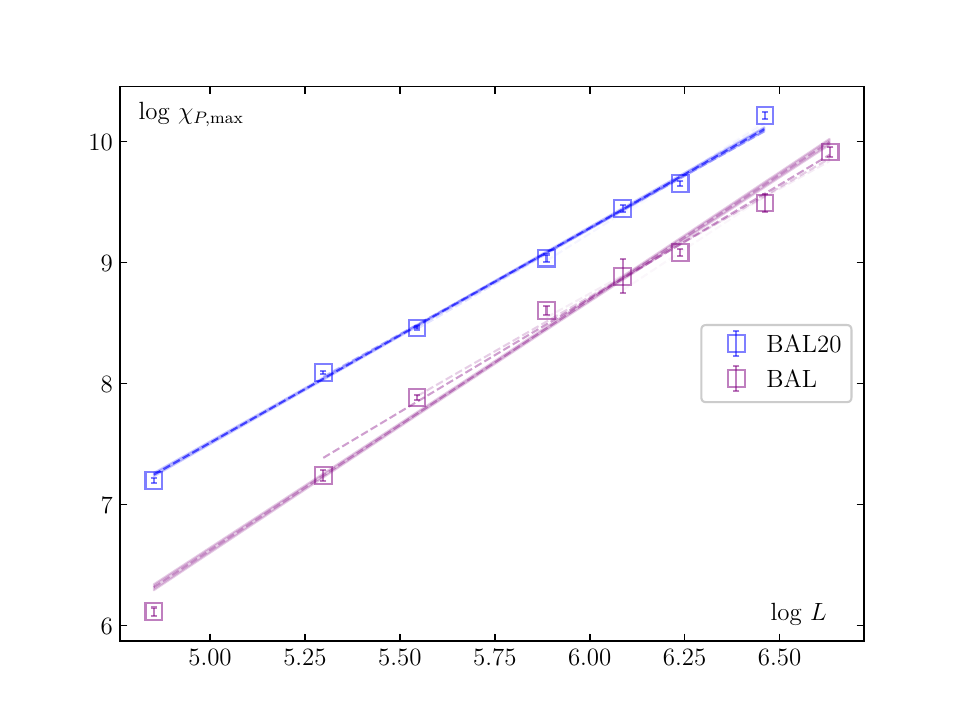}
\caption{Extraction of $\Tc$ and critical exponents for the 2-$d$ Ising model 
using the CNN output layer. {\it Left}: Extraction of $T_c$ and $\nu$. 
The grey, dotted line indicates the theoretical value.
{\it Right}: Extraction of $\gamma$. Bands show a linear fit to all data with error.
Dashed lines indicate linear fits neglecting the smallest lattices, which
enter the BMA. The transparency of each line indicates $\pr{M_n|D}$ with darker 
lines indicating greater model weight.}
\label{fig:exponents}
\end{figure*}

\section{Results and deep learning model performance}\label{sec:results}

In \figref{fig:PwithL} we show the average output layer for each lattice size
for the BAL20 (left) and BAL (right) models. One sees a substantial difference
in performance when comparing lattice sizes; in particular $\ev{P}$ for the
BAL20 model becomes a gradually steeper function of $T$ as the thermodynamic
limit is approached, which is qualitatively the expected behavior for
an order parameter. On the other hand, for the BAL model this behavior is only
partially respected by the largest and smallest volumes. The system with $L=256$ in particular
departs quite substantially from the other curves, with $\ev{P}$ just
barely exceeding 0.6 at $T=2.25$.

Under the assumption that $\ev{P}$ behaves as an effective order parameter,
we estimate pseudocritical temperatures $\Tc(L)$ using the peak location
of the prediction susceptibility $\chi_P$ for each $L$. We reweight $\chi_P$
according to Eq.~\eqref{eq:RW} to find the peak. 
The peak location found by reweighting depends on the starting point
$\beta$ used in Eq.~\eqref{eq:RW}. We reweight using points closest
to the maximum, then estimate a systematic uncertainty equal
to half the spread of the $\Tc(L)$ found for each $\beta$. The total error
is computed by adding the statistical and systematic uncertainties
in quadrature. 

This process is illustrated for each CNN model using
the largest lattices in \figref{fig:RW} (top). The range in $\beta$
for which reweighting is reliable decreases exponentially with $V$;
that the reweighting curves agree on our largest volumes shows
that our reweighting procedure is well controlled.
In \figref{fig:RW} (bottom) we show the corresponding reweighted
average predictions. For these volumes, we find $\Tc(L)$ to be at or close
to the region where $\ev{P}=0.5$

Armed with estimates for $\Tc(L)$ and $\chimax$, we can extract critical exponents.
In \figref{fig:exponents} we show results for the BAL20 and BAL models.
We begin by extracting $\Tc$ and $\nu$ by carrying out a fit using
Eq.~\eqref{eq:nuTc}. To improve numerical stability, we recast this equation as
\begin{equation}\label{eq:fit}
\log L=-\nu\log |\Tc(L)-\Tc|.
\end{equation}
Since our uncertainties are in the independent variable $T$, we carry out this
fit using orthogonal distance regression\footnote{We do our orthogonal distance
regression using the software of 
\textsc{SciPy}~\cite{Virtanen:2019joe}. These fits are cross-checked using
\textsc{gvar}~\cite{gvarGitHub}.}. Results of the fits are in
\figref{fig:exponents} (left). As was reported in Ref.~\cite{Bachtis:2020dmf},
the results for the BAL20 model are quite good. We find
$\Tc^{\rm BAL20}=2.26828(61)$ and 
$\nu^{\rm BAL20}=1.017(41)$
with a residual variance of 1.547, which is in excellent
agreement with the analytic result.

We found the finite-size scaling
of the BAL model not to be described as well by the fit of Eq.~\eqref{eq:fit};
in particular the fit has a residual variance of 28.982.
This is due to the outlier at $L=256$ along with the fact
that the BAL model seems to find the same $\Tc(L)$ within error
for $L=360,$ 440, and 512. That $\Tc(256)$ is quite low is perhaps not
surprising in light of \figref{fig:PwithL} (right), where one sees
that the BAL model had substantial difficulty to classify the ferromagnetic
phase for $L=256$.
This behavior may be a result of overfitting,
as we attempted to follow the architecture of Ref.~\cite{Bachtis:2020dmf} exactly. This architecture
uses no regularization. 
Still, our fit yields
$\Tc^{\rm BAL}=2.2664(29)$ and
$\nu^{\rm BAL}=1.21(12)$,
which are both statistically compatible with the analytic results.
We note that the relative uncertainty for $\Tc$ is roughly 4.76
times greater for the BAL model than the BAL20 model, and similarly,
the relative uncertainty for $\nu$ is roughly 2.46 times larger.

Next we show the extraction of $\gamma$. Following \equatref{eq:gamma},
we carry out a two-parameter, linear fit
\begin{equation}
\log \chimax=C+\frac{\gamma}{\nu}\log L.
\end{equation}
using conventional least-squares methods. Results are shown in
\figref{fig:exponents} (right). The BAL20 fit gives
$(\gamma/\nu)^{\rm BAL20}=1.776(17)$
with a $\chidof=8.60$, in agreement with the known result.
By contrast the BAL model seems unable to capture $\gamma$.
We find
$(\gamma/\nu)^{\rm BAL}=2.069(24)$
with a $\chidof=18.25$. While this number is not statistically compatible with
the known result, $\chimax$ for this CNN behaves qualitatively reasonably:
We see that $\log\chimax$ scales very roughly linearly with $\log L$ with a slope
roughly 18\% larger than the analytic value.

The singular behavior expressed by \eqref{eq:nuTc} and \eqref{eq:gamma} is
a good description for large $L$. As $L$ decreases, regular contributions become
more important, which may lead to significant departures from these scaling laws.
The predictions from the different CNN models $P^{\rm BAL20}$
and $P^{\rm BAL}$ represent different observables with different regular contributions, 
and hence there is no guarantee
that the relation \eqref{eq:gamma} will be a good approximation 
starting with the same minimum $L$ for both quantities. 
To understand the extent to which the inability to capture $\gamma/\nu$ can be
explained by a higher threshold beyond which \eqref{eq:gamma} becomes a
suitable description,
we try fits that leave out progressively larger $L$ values among the smallest lattices.

\begin{table}
\centering
\caption{\label{tab:gammaBMA}
Results of the fits shown in \figref{fig:exponents} (right)
along with their model weights for each $M_n$, i.e. for each choice
of $L_{\rm min}$. Model weights are rounded to the second digit.} 
\begin{ruledtabular}\begin{tabular}{llclr}
 & BAL20 & & BAL & \\
$L_{\rm min}$ & $\left(\gamma/\nu\right)_n$ & $\pr{M_n\given D}$
& $\left(\gamma/\nu\right)_n$ & $\pr{M_n\given D}$\\ 
\hline
128 &1.776(17) &0.61 &2.069(24) &0.32 \\
200 &1.745(20) &0.23 &1.878(32) &0.38 \\
256 &1.815(27) &0.10 &1.807(37) &0.19 \\
360 &1.977(71) &0.04 &1.733(69) &0.07 \\
440 &2.09(12)  &0.02 &2.07(13)  &0.03 \\
\end{tabular}\end{ruledtabular}
\end{table}

Each combination of analysis choices, in this case the fit of Eq.~\eqref{eq:gamma} 
equipped with choosing a number of $L$ data to prune, defines a statistical 
model\footnote{We have tried to distinguish the
use of the word ``model" in this sense from deep-learning models by calling the latter
``deep learning model", ``BAL model", or ``BAL20 model".} $M_n$.
To compute an average from this process we employ Bayesian model averaging 
(BMA)~\cite{Jay:2020jkz,Neil:2022joj,Neil:2023pgt},
\begin{equation}\label{eq:BMAMean}
    \ev{\gamma/\nu}_{\rm BMA} = \sum_{n=1}^{N_M}\ev{\gamma/\nu}_{n} \pr{M_n \given D},
\end{equation}
where the sum runs over models $M_n$, $N_M$ is the number of models, and
$\ev{\gamma/\nu}_n$ is the result for $\gamma/\nu$ found in model $M_n$.
The model weight, that is the probability of the model $M_n$ given the data $D$, 
is given by~\cite{Neil:2023pgt}
\begin{equation}\label{eq:modelProb}
    \pr{M_n \given D} = \pr{M_n} \exp \left[-\frac{1}{2}
    \left(\chi^2_n
    +2 k+N_{\rm cut}\right)\right], 
\end{equation}
where $\pr{M_n}$ is the prior probability of model $M_n$, $\chi^2_n$ 
gives\footnote{More precisely this is $\chi^2_{\rm data}$, the $\chi^2$
computed not taking into account any priors on fit parameters. We use no priors for fit parameters
in this study, so we do not make the distinction.} 
$\chi^2$ evaluated for model $M_n$, $k$ is the
number of fit parameters, and $N_{\rm cut}$ is the number of sizes $L$ trimmed
from the data set. This choice of model weight corresponds to using the
Bayesian Akaike information criterion of Ref.~\cite{Neil:2023pgt}.
The model weight penalizes poor fits, large numbers
of fit parameters, and cutting away data. For our
case, $k=2$, and we take $\pr{M_n}=1/N_M$, a choice which gives no
{\it a priori} preference for one model over another. The variance in
the BMA average $\ev{\gamma/\nu}$ is computed through~\cite{Neil:2023pgt}
\begin{equation}\begin{aligned}\label{eq:BMAVar}
\sigma^2_{\rm BMA} &= \sum_{n=1}^{N_M} \sigma_{n}^{2} \pr{M_{n} \given D}\\
&~~~~+\sum_{n=1}^{N_M}\ev{\gamma/\nu}_{n}^{2} \pr{M_{n} \given D}
-\ev{\gamma/\nu}_{\rm BMA}^2,
\end{aligned}\end{equation}
where $\sigma_n^2$ is the variance in $\ev{\gamma/\nu}_n$.
This can be interpreted as a statistical uncertainty combined with
a systematic measure of  the model spread, given by the second and third terms.

The fits for each $M_n$ entering the BMA are indicated by dashed lines 
in \figref{fig:exponents} (right),
where the transparency of the dashed line indicates $\pr{M_n\given D}$.
The exact fit results and model weights are reported in
\tabref{tab:gammaBMA}.
With the exception of $L_{\rm min}=440$, increasing the minimum $L$ 
used decreases the slope for the BAL model. The model weight is at
its largest at $L_{\rm min}=200$ and decreases monotonically thereafter,
being penalized by $N_{\rm cut}$ in Eq.~(\ref{eq:modelProb}). That the exclusion
of the smallest lattice rewards the model probability more than $N_{\rm cut}$ 
punishes it may be interpreted as evidence that regular contributions
for this observable become small starting around $L_{\rm min}=200$.
The BMA yields
$(\gamma/\nu)^{\rm BAL}_{\rm BMA}=1.92(13)$.
The central value decreases compared to the result
using no BMA. It remains
high compared to the analytic result, but is now compatible within the
substantially larger uncertainty. 
For comparison, we also carry out a BMA
for the BAL20 model. Again the fits are indicated in the figure using dashed lines.
In contrast to the BAL model, the fits are much more closely aligned; the fits
with non-negligible model weight overlap with the original fit. Moreover
we find no benefit to excluding the smallest lattice, with model weights
monotonically decreasing with increasing $L_{\rm min}$ throughout. We find
$(\gamma/\nu)^{\rm BAL20}_{\rm BMA}=1.786(65)$.
As was the case with $\Tc$ and $\nu$, the BAL model has
a larger relative uncertainty, in this case roughly
a factor 1.86.
Our final results for critical parameters are summarized in \tabref{tab:finalResults}.

\begin{table}
\centering
\caption{\label{tab:finalResults}
Critical parameters of each ML model compared against
the analytic result~\cite{Onsager:1943jn}.}
\begin{ruledtabular}\begin{tabular}{lllr}
& Onsager & BAL20 & BAL \\ 
\hline
$\Tc$        & 2.269185 & 2.26828(61) & 2.2664(29)\\
$\nu$        & 1        & 1.017(41)   & 1.21(12)\\
$\gamma/\nu$ & 1.75     & 1.786(65)   & 1.92(13) \\
\end{tabular}\end{ruledtabular}
\end{table}

CNNs trained only at $T=0$ and $T=\infty$ like the BAL model may in
general struggle to accurately extract $\gamma$ compared to CNNs trained with
many examples. To gain some understanding whether this approach contains any inherent
obstacles to determining $\gamma$, consider
a configuration $C$ just above $\Tc(L)$. Take a handful of its down spins and
flip them to create a new configuration $C'$. The BAL20 model should more readily 
recognize that $C'$ is more ferromagnetic than $C$ compared to the BAL model, which
is shown no intermediate temperature examples. Hence BAL20 can assign appreciably 
different $P$s to $C$ and $C'$. At the same time, the energy cost of transforming 
from $C$ to $C'$ is low, and correspondingly $C$ and $C'$ are generated with 
roughly similar probabilities at $\Tc$. This will lead to greater variance in 
$\ev{P^{\rm BAL20}}$ than $\ev{P^{\rm BAL}}$ at $\Tc(L)$. This is reflected in
\figref{fig:exponents} (right): One sees that the maximum of the prediction
susceptibility, i.e. the variance in the prediction at $\Tc(L)$, 
is larger for the BAL20 model than the BAL model for
all lattice sizes. 
If $\log\chimax^{\rm BAL20}(L)>\log\chimax^{\rm BAL}(L)$ for all $L$, then
in particular these functions are not equal, so there is no
guarantee that the slopes will match.

On a related note, the NN taken from Ref.~\cite{Bachtis:2020dmf} was
optimized to 20 example temperatures. Applying this NN to two temperature examples
only, as in the BAL model, means that the NN sees an order of magnitude fewer data
during training. This disparity in sample size likely impacts the performance of BAL
as well.

When exploring the finite-size scaling of a system, 
the critical exponent $\beta$ can be extracted from the 
finite-size scaling of the magnetization
$m$ via
\begin{equation}\label{eq:beta}
    \ev{m}\big|_{\Tc(L)}\sim L^{-\beta/\nu}.
\end{equation}
Although in many ways $\ev{P}$ behaves similarly as $\ev{m}$, 
this pseudo order parameter cannot exhibit this same scaling 
form~\eqref{eq:beta}.
In particular since the pseudocritical
temperature $\Tc(L)$ is the point at the phase boundary,
configurations at this temperature ought to be equally often
classified as ferromagnetic and paramagnetic, and hence
one expects $\ev{P\left(\Tc(L)\right)}\approx1/2$,
i.e. it is a constant independent of $L$.
As mentioned earlier, we see in \figref{fig:RW} (bottom)
that $\ev{P}$ at $\Tc(L)$ is compatible with 0.5
within error for the largest lattices of both deep learning
models.

\section{Summary and outlook}\label{sec:summary}

In this paper, we studied the capabilities of a CNN (BAL model) using a minimal set of examples,
specifically examples at $T=0$ and $T=\infty$, comparing its performance to the 
same CNN trained on 20 temperatures near $\Tc$ (BAL20 model). 
We confirm the findings of Ref.~\cite{Bachtis:2020dmf} that 
the output layer can be reweighted and that
the BAL20 model easily extracts $\Tc$, $\nu$, and $\gamma$.  
The BAL model 
results for
$\Tc$ and $\nu$ are statistically compatible with literature results, 
albeit with poor fit quality and correspondingly higher
uncertainty. A BMA of fits for $\gamma$ for the BAL model prefers fits
over our largest volumes. This behavior may be due in part to regular contributions
remaining substantial at larger $L$ than the BAL20 model and in part to the performance
of the BAL model itself. The larger uncertainties, lower fit qualities, and relative
difficulty extracting $\gamma$ exposes to some extent the role of examples
at intermediate temperatures.
Our findings that our minimally trained network delivers $\Tc$ accurately
with a relative uncertainty of roughly 1\%, along with an estimate for $\nu$
that is compatible with the Onsager solution and a $\gamma/\nu$ whose BMA
is compatible,
are broadly concordant with studies that show
minimal training regimens can identify the phase transition order
and the general vicinity of $\Tc$ for Potts 
models~\cite{Li:2017xaz,Tan:2019eih} and with a study showing
that two examples near the $g=0.7$ $\Tc$ of the frustrated 2-$d$
Ising model are sufficient to estimate $\Tc$ at 
$g=0.8$~\cite{Li:2025rbf}. As already noted in Refs.~\cite{Li:2017xaz,Tan:2019eih},
reducing the training set size also has the advantage of 
significantly reducing the computational cost of training.

In this study, we looked at only one architecture to focus on the effect the choice
of examples has on a successful extraction of critical parameters. 
This architecture was empirically optimized in Ref.~\cite{Bachtis:2020dmf} 
for their regimen of 20 example temperatures. Therefore it is plausible that
a different architecture would be better suited for having two examples only.
For example, introducing regularization may reduce overfitting: one might
consider dropout regularization, which can have lower test classification
error than $L_2$ regularization~\cite{dropout},
or weight decay regularization, which substantially improves Adam's
generalization performance on image classification datasets~\cite{weightDecay}.
One could also explore using batch normalization, which can act as a regularization,
is well suited to CNNs, and
allows one to use much higher learning rates and be less careful about 
initialization~\cite{batch}.
Further architectures and classical statistical physics models 
will be explored in future work.

We find our results encouraging for eventual application of supervised learning to
systems with crossovers. In particular, the ability of a minimally trained model to
find not just $\Tc$ and $\nu$ but also to some degree $\gamma$ 
indicates that the model still learns
salient features of configurations that flag a change of phase.
Thinking about systems with a crossover, like the QCD crossover, 
one may imagine training at very high temperatures
and very low temperatures, 
using this approach as a first step that identifies a range of promising
pseudocritical temperatures. This range could then be refined
for example by
using multiple temperatures where all known physical
observables agree on the phase.
\vspace{.2cm}
\section*{Acknowledgments}
We acknowledge fruitful discussions with Gregory Morrison, Kevin Bassler, and Ricardo Vilalta. The research reported in this work made use of computing 
facilities of the USQCD Collaboration, which are funded by the Office of Science
of the U.S. Department of Energy.
DAC was supported in part by the U.S. Department of Energy, Office of Science,
under the Funding Opportunity Announcement Scientific Discovery through Advanced 
Computing: High Energy Physics, LAB 22-2580 and by the National Science Foundation 
under Grant No. PHY23-10571.
The authors acknowledge the use of the Carya Cluster and the advanced support from the Research Computing Data Core at the University of Houston to carry out the research presented here.
This material is based upon work supported by the U.S. National Science Foundation under grants No. PHY-2208724 and PHY-2116686, and within the framework of the MUSES collaboration, under Grant  No. OAC-2103680. This material is also based upon work supported by the U.S. Department of Energy, Office of Science, Office of Nuclear Physics, under Award Number DE-SC0022023, as well as by the National Aeronautics and Space Agency (NASA) under Award Number 80NSSC24K0767. MHJ was supported in part by the U.S. Department of Energy under award number DE-SC0024586 and the U.S. National Science Foundation (NSF) under grant No. PHY-2310020.

\bibliography{bibliography}
\end{document}